# TRANSFERRED THIN FILM LITHIUM NIOBATE AS MILLIMETER WAVE ACOUSTIC FILTER PLATFORMS

*Omar Barrera[1], Sinwoo Cho[1], Kenny Hyunh[2], Jack Kramer[1], Michael Liao[2]*
*Vakhtang Chulukhadze[1], Lezli Matto[2], Mark S. Goorsky[2], and Ruochen Lu[1]*
[1]University of Texas at Austin, and [2]University of California Los Angeles

## ABSTRACT

This paper reports the first high-performance acoustic filters toward millimeter wave (mmWave) bands using transferred single-crystal thin film lithium niobate ($LiNbO_3$). By transferring $LiNbO_3$ on the top of silicon (Si) and sapphire ($Al_2O_3$) substrates with an intermediate amorphous Si (aSi) bonding and sacrificial layer, we demonstrate compact acoustic filters with record-breaking performance beyond 20 GHz. In the LN-aSi-$Al_2O_3$ platform, the third-order ladder filter exhibits low insertion loss (IL) of 1.62 dB and 3-dB fractional bandwidth (FBW) of 19.8% at 22.1 GHz, while in the LN-aSi-Si platform, the filter shows low IL of 2.38 dB and FBW of 18.2% at 23.5 GHz. Material analysis validates the great crystalline quality of the stacks. The high-resolution x-ray diffraction (HRXRD) shows full width half maximum (FWHM) of 53" for $Al_2O_3$ and 206" for Si, both remarkably low compared to piezoelectric thin films of similar thickness. The reported results bring the state-of-the-art (SoA) of compact acoustic filters to much higher frequencies, and highlight transferred $LiNbO_3$ as promising platforms for mmWave filters in future wireless front ends.

## INTRODUCTION

Mobile communication relies heavily on acoustic wave resonators for compact front-end filters in sub-6 GHz frequency bands [1]. Commercially successful acoustic filters can be mostly classified into two groups: surface acoustic wave (SAW) and bulk acoustic wave (BAW) devices [2], [3]. Both solutions share the same advantages of a small footprint compared to the electromagnetic (EM) counterparts, thanks to the short wavelengths of acoustics, suitable for handheld devices [4], [5]. As the demand for higher data rates increases, wireless communication is exploring millimeter wave (mmWave) frequency bands. Consequently, there is a growing need for compact filters at mmWave with low loss and great frequency selectivity [6], [7]. On the one hand, frequency-scaling acoustic filters have suffered excessive loss [Fig. 1 (a)] and bandwidth reduction [Fig. 1 (b)] due to a lack of high-performance mmWave acoustic resonator platforms [8]–[12]. On the other hand, circuit and EM solutions have disadvantages, such as the low $Q$ of lumped components and the bulkiness of enclosures in cavity filters [13]. This work focuses on developing acoustic technologies for low-loss filter platforms at mmWave.

Increasing the operation frequencies of acoustic resonators is simple in principle, but maintaining device performance is challenging [14]. More specifically, achieving simultaneously high quality factor ($Q$) and electromechanical coupling ($k^2$) at mmWave is difficult [15]–[22]. For SAW, reducing the electrode width translates to shorter wavelengths and higher frequencies, but this method is contingent on feature size (sub-100 nm) and thus limited by lithography [23], [24]. Besides, the routing resistance and self-inductance of interdigitated electrodes exacerbate for narrower electrodes. For BAW, the frequency is inversely proportional to the piezoelectric layer's thickness, while at mmWave, the film stack becomes significantly thin (sub-100 nm), causing challenges in maintaining piezoelectric film quality and lowering electrical loss [25]. Additionally, the resonators required to realize a 50Ω matched filter must be very small to obtain the low values of static capacitance ($C_0$) required for the network, which inevitably results in lossy devices [6], [26]. Thus, a new piezoelectric acoustic platform at mmWave is needed.

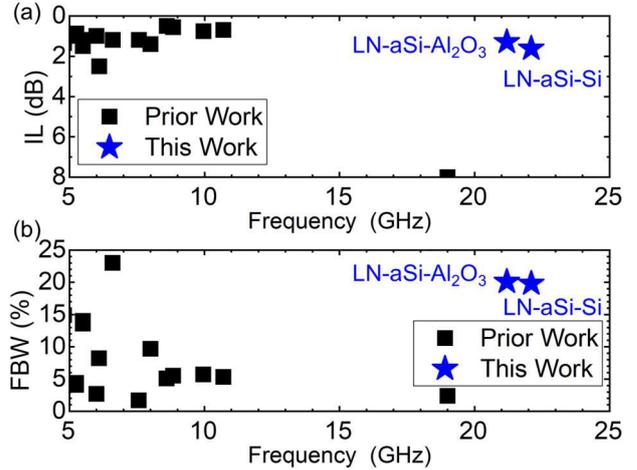

*Fig. 1. Surveys indicating (a) IL and (b) FBW for reported acoustic filters with $f_c$ beyond 5 GHz.*

Lately, alternate methods of exciting bulk acoustic waves in the thin film have resulted in laterally excited resonators on lithium niobate (LiNbO3) operating in thickness-shear modes [27], [28]. Specifically, a new proposed stack involving transferred thin-film $LiNbO_3$ on low-loss substrates with an intermediate amorphous silicon layer has recorded resonators with impressive Q and $k^2$ [29], [30]. In this work, we exploit this material stack on 2 different carrier substrates, silicon (Si) and sapphire ($Al_2O_3$), demonstrating the capability of transferred film $LiNbO_3$ as a potential platform for mmWave acoustic filters. The fabricated devices exhibit low IL of 1.62 dB and 2.38 dB, with large FBW of 19.8% and 18.2% for the Si and $Al_2O_3$ substrates, respectively, elevating the current state of the art (Fig. 1). The third-order ladder filters are well matched to 50Ω in a compact footprint of 0.56mm$^2$.

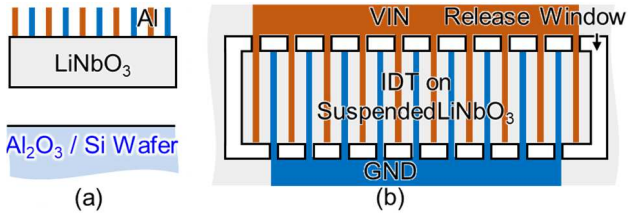

*Fig. 2. (a) Cross-sectional and (b) top resonator schematics.*

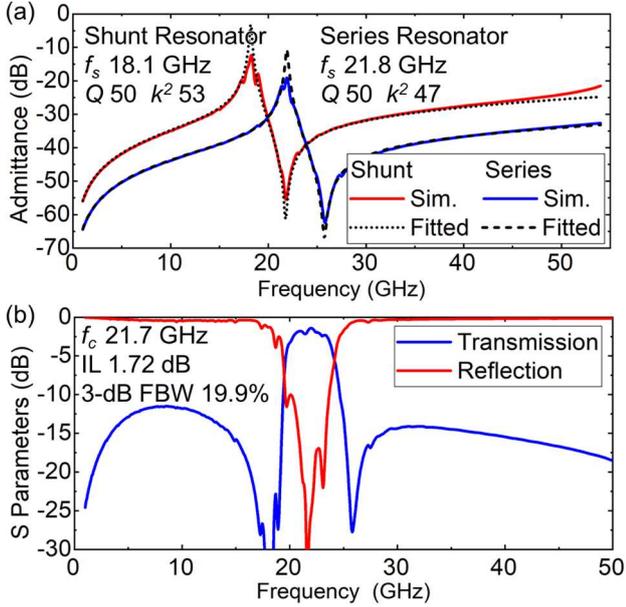

*Fig. 3 (a) Simulated admittance, inset key specifications and vibration mode shape. (b) Simulated filter response.*

These prototypes showcase a proof of concept as potential candidates for mmWave acoustic filters.

## DESIGN AND SIMULATION

The cross-sectional and top views of the design are illustrated in Fig. 2. The active region consists of interdigitated transducers (IDT) on a thin-film LiNbO$_3$ piezoelectric layer. Thickness-shear waves, i.e., first-order antisymmetric (A1) mode, are excited via the $e_{15}$ piezoelectric coefficient by the electrodes. Aluminum is selected for the metal layer with a thickness of 350 μm, and the spacing between electrodes is 3 μm, following the guidelines reported in [29]. The design approach is identical for both Si and Al$_2$O$_3$ carrier substrates, as the resonators are released and only connected with the substrate with anchors. However, the difference between the two stacks lies in the EM feedthrough and transfer film quality (Figs. 4-5), which will be discussed later.

The resonator structure is validated using COMSOL finite element analysis, and the results are plotted in Fig. 3 (a). The results show a frequency shift, which is necessary for filter synthesis, where the parallel resonance ($f_p$) of the shunt resonator and the series resonance ($f_s$) of the series resonator are intended to overlap. This shift is achieved using different thicknesses of LiNbO$_3$ for the shunt and series resonators respectively. $Q$ is set as 50 based on previously measured data, and $k^2$ is extracted as $k^2 = \pi^2/8 \cdot (f_p^2/f_s^2 - 1)$ from a fitting using the Butterworth-Van Dyke (MBVD) model [31]. The values of $C_0$ are selected to obtain minimum IL in a 50 Ω third-order ladder filter.

The obtained results are exported into a circuit simulator to assess the expected performance of the filter. The results [Fig. 3 (b)] show a filter centered at 21.7 GHz with IL of 1.72 dB and a 19.9% FBW, promising performance at this frequency range.

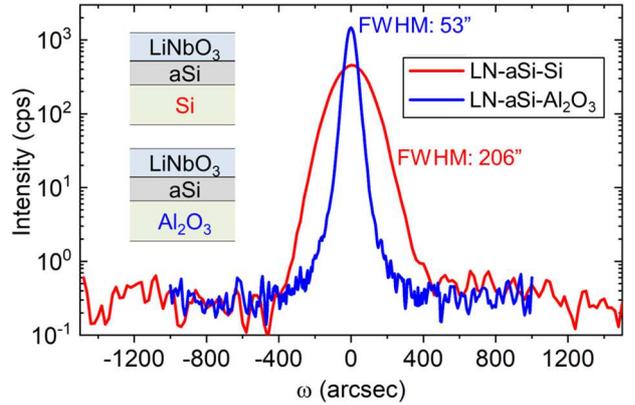

*Fig. 4. Measured HRXRD rock curves of the 128° Y-cut LiNbO$_3$ crystal in both platforms show high crystal quality.*

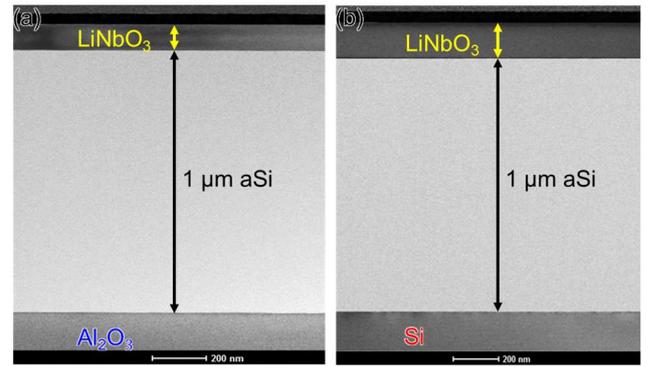

*Fig. 5. Stack information measured with TEM pictures of (a) LN-aSi-Al$_2$O$_3$ and (b) LN-aSi-Si stacks.*

## FABRICATION AND MEASUREMENT

The wafers used for this work were provided by NGK Insulators Ltd. The two proposed stacks (LN-aSi-Si & LN-aSi-Al$_2$O$_3$) are analyzed using material characterization techniques. First, rocking curves are measured using high-resolution x-ray diffraction (HRXRD) to validate the crystal quality of the material stacks (Fig. 4). The results show full width at half maximum (FWHM) of 53" for Al$_2$O$_3$ and 206" for Si, comparable to that reported in other thin film LiNbO$_3$ layers. Next, side profile images of the stack are measured using transmission electron microscopy (TEM). The results show clean interfaces between material boundaries, suggesting good crystallinity of the materials (Fig. 5). The images also reveal a slightly thicker LiNbO$_3$ layer for the LN-aSi-Si compared to the LN-aSi-Al$_2$O$_3$ stack, which suggests small thickness variations across the wafers.

The fabrication starts by trimming the LiNbO$_3$ thickness of the samples down to 90 nm, which provides the base start thickness used for the shunt resonators. The trimming is done using ion milling; in [27], the authors have validated good preservation of the crystal with this approach. Afterward, etching windows are patterned using lithography and etched into the aSi layer. The long lateral etch windows divide the devices into resonator banks, expediting the release process. Next, local regions for placement of the series resonators are defined using lithography. Afterward, a second round of ion milling is used on these exposed local trimming regions for a target thickness of 75 nm. The step height difference between these regions is accurately monitored using atomic force microscopy (AFM), with a measured height of 15±1 nm. EBL lithography is used to pattern the fine features of the metal layer, and Al is evaporated for the metal deposition. The devices are then released by selectively etching the aSi intermediate layer using xenon difluoride (XeF$_2$). Figs. 6 (a)-(c) show optical images of the standalone resonators as well as the filter for the LN-aSi-Al2O3 sample and the LN-aSi-Si filter in Fig. 6(d).

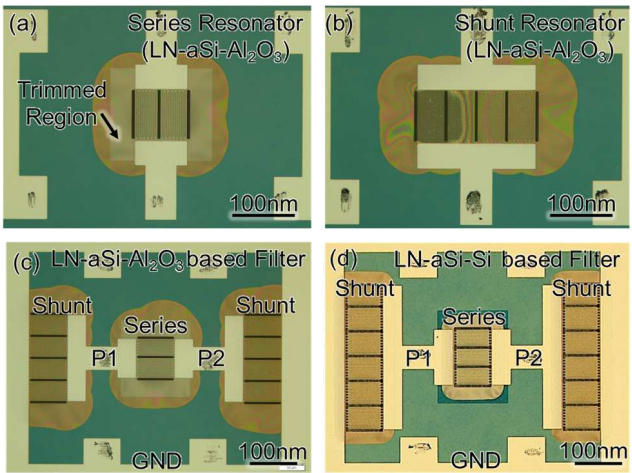

Fig. 6. Fabricated (a) series, (b) shunt resonator, and (c) filter in LN-aSi-Al$_2$O$_3$. (d) Filter in LN-aSi-Si.

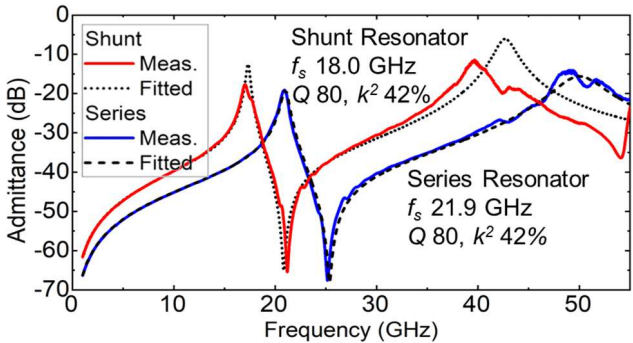

Fig. 7. Measured series & shunt resonator in LN-aSi-Al$_2$O$_3$.

The standalone resonators are measured at room temperature using a 67 GHz Keysight VNA. Fig. 7 shows the admittance response for the resonators on LN-aSi-Al$_2$O$_3$. The series and shunt devices exhibit nearly identical figures of merit, with $k^2$ of 42% and $Q$ of 80. $k^2$ is extracted using MBVD and considering the series routing resistance and inductance observed in IDT devices. A good overlap between the $f_p$ of the shunt resonator and the $f_s$ of the shunt resonator can be observed around 22 GHz, corroborating the accuracy of the local trim.

The filter response for LN-aSi-Al$_2$O$_3$ (Fig. 8a) shows IL of 1.62 dB, FBW of 19.8% at 22.1 GHZ $f_c$, whereas the LN-aSi-Si (Fig. 8b) sample has 2.38 dB IL with 18.2% FBW at $f_c$ of 23.5 GHz. The measurements for filters and standalone resonators are in good agreement with the simulations, the discrepancy in $f_c$ between the 2 filters is due to the small variations in the thickness of the LiNbO$_3$ layer across the sample surface. These results show remarkable frequency scaling and improved FBW compared to the SoA (Fig. 1) while maintaining low IL. The out-of-band characteristics are similar to other works using 3 resonator ladder filters [8], [32].

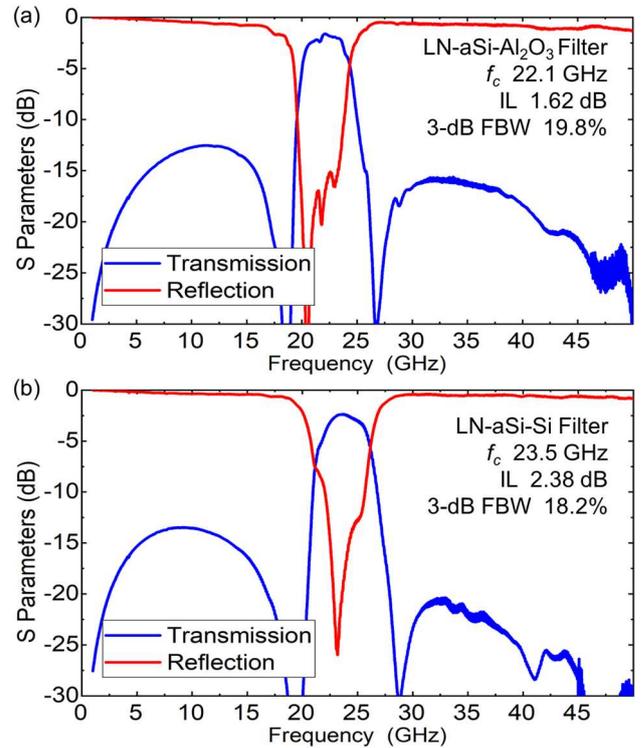

Fig. 8. Measured transmission and reflection of fabricated filters in (a) LN-aSi-Al$_2$O$_3$, and (b) LN-aSi-Si.

## CONCLUSION

This work presents the suitability of transferred thin-film LiNbO$_3$ as a platform for mmWave filters. The prototypes show promising low-loss and wideband response, well above the current SoA. These unprecedented results are backed by material analysis, which shows excellent preservation of the piezoelectric layer crystal quality. The reported devices demonstrate the potential of transfer thin-film LiNbO$_3$ platforms as prospective candidates for the next generation of mmWave acoustic filters.

## ACKNOWLEDGMENT

The authors thank the DARPA COFFEE program for funding support and Dr. Ben Griffin for helpful discussions.

**CONTACT**

Omar Barrera, omarb@utexas.edu